\documentclass[twocolumn,amsmath,amssymb,showpacs,superscriptaddress,pra]{revtex4}
\usepackage{graphicx}% Include figure files
\usepackage{dcolumn}% Align table columns on decimal point
\usepackage{bm}% bold math

\newcommand{\bs}{\boldsymbol}
\newcommand{\mb}{\mathbf}

\begin{document}

%\preprint{APS/123-QED}

\title{Spin quantum correlations of relativistic particles}

\author{Pablo L. Saldanha}\email{saldanha@df.ufpe.br}
\affiliation{Department of Physics, University of Oxford, Clarendon Laboratory, Oxford, OX1 3PU, United Kingdom}
\affiliation{Departamento de F\'isica, Universidade Federal de Pernambuco, 50670-901, Recife, PE, Brazil}
\author{Vlatko Vedral}
\affiliation{Department of Physics, University of Oxford, Clarendon Laboratory, Oxford, OX1 3PU, United Kingdom}
\affiliation{Centre for Quantum Technologies, National University of Singapore, Singapore}
\affiliation{Department of Physics, National University of Singapore, Singapore}
%\affiliation{


\begin{abstract} 
We show that a pair of massive relativistic spin-1/2 particles prepared in a maximally entangled spin state in general is not capable of maximally violating the Clauser-Horne-Shimony-Holt (CHSH) version of Bell's inequalities without a post-selection of the particles momenta, representing a major difference in relation to non-relativistic systems. This occurs because the quantization axis of the measurements performed on each particle depends on the particle velocity, such that it is not possible to define a reduced density matrix for the particles spin. We also show that the amount of violation of the CHSH inequality depends on the reference frame, and that in some frames the inequality may not be violated.
\end{abstract}

                             % Classification Scheme.%\keywords{Suggested keywords}%Use showkeys class option if keyword
  % 03.65.Ud 	Entanglement and quantum nonlocality 
  % 03.65.Ta 	Foundations of quantum mechanics; measurement theory   
  % 03.30.+p 	Special relativity
  % 03.67.-a 	Quantum information
                              
\maketitle

Special relativity and quantum mechanics were two theories constructed in the last century that completely changed our way to see nature, being the foundations of present-day theoretical physics. One of the most striking features of quantum mechanics is entanglement, that leads to quantum correlations among parties of a system that are stronger than what is allowed by classical physics \cite{bell,peres,clauser69}. Recently questions regarding the behaviour of the entropy and entanglement of quantum systems in different reference frames gave rise to the field of relativistic quantum information \cite{czachor97,peres02,peres04}. % that tries to give a better understanding of the union between special relativity and quantum mechanics. 
 Since then, many studies of relativistic effects on spin quantum correlations of massive particles  have appeared in the literature \cite{czachor97,alsing02,gingrich02,ahn03,li03,terashima03,lee04,kim05,caban05,lamata06,jordan07,landulfo09,caban10,friis10,choi11}.

In the seminal work of Peres, Scudo and Terno \cite{peres02}, they showed that the reduced density matrix for the spin of a relativistic particle, that should give ``the statistical predictions for the results of measurements of spin components by an ideal apparatus which is not affected by the momentum of the particle'' \cite{peres02}, is not covariant under Lorentz transformations. This occurs because under a Lorentz boost the particle spin undergoes a Wigner rotation \cite{wigner39,weinberg}, which correspond to a momentum-dependent change of the particle spin state. 
%This phenomenon may generate correlations between the spin and the momentum components of the particle state in some reference frames when there were no correlations in other frames, and these correlations increase the entropy of the reduced spin state.
 One aspect extensively studied in the posterior works on relativistic quantum information was the influence of the Wigner rotations  on the amount of entanglement of the reduced spin density matrix of a system with two ore more particles in different reference frames \cite{gingrich02,li03,lamata06,jordan07,landulfo09,friis10}. However, in our recent work \cite{saldanha12} we discussed that since it is not possible to measure the spin of a relativistic particle in a independent way from its momentum, the definition of a reduced density matrix for the particle spin is meaningless. The ideal apparatus which is not affected by the particle momentum conjectured by Peres \textit{et al.} \cite{peres02} does not exist, contradicting the assumptions (explicitly or implicitly) made by the cited works \cite{peres02,peres04,gingrich02,li03,lamata06,jordan07,landulfo09,friis10,choi11}.

A second aspect studied in many of the previous works on the subject is the influence of the dependence of the Pauli-Lubanski (or similar) spin operators with the particles momenta on the amount of violation of Bell's inequalities with relativistic particles \cite{czachor97,ahn03,lee04,kim05,caban05,caban10,friis10}. But in our recent work \cite{saldanha12} we discussed that to use the Pauli-Lubanski (or similar) spin operators to describe spin measurements, spin must couple to a quantity that transforms as part of a 4-vector under Lorentz transformations in the measuring apparatus. But we do not know if such a coupling exists in nature. If spin couples to an electromagnetic field in the measuring apparatus, like in the Stern-Gerlach measurements, the spin operators must transform as part of a tensor under Lorentz transformations to guarantee the invariance of the outcomes probabilities, with different predictions in relation to the Pauli-Lubanski treatment \cite{saldanha12}.  

For the reasons described in the previous paragraphs, we believe that the analysis of spin quantum correlations of relativistic systems must be revisited. Here we apply our method to the case of two entangled spin-1/2 massive particles to show how the maximum amount of violation of the Clauser-Horne-Shimony-Holt (CHSH) version of Bell's inequalities \cite{clauser69} depends on the velocity distribution of the particles. This represents a major difference in relation to non-relativistic systems, and was not considered in the previous works on the subject. We also show that observers in different reference frames may obtain different amounts of violation for the CHSH inequality, and that some of them may not be able to violate this inequality without a post-selection of the particles momenta. 
%In a relativistic scenario, a spin entangled system may not be able to present non-classical spin correlations. 

In the present work, we will consider the following state for two spin-1/2 particles labelled by $a$ and $b$ in the center-of-mass rest frame:
\begin{equation}\label{singlet}
	|\Psi\rangle=\int d^3p\, \psi(\mb{p})\Big[|\mb{p},\uparrow\rangle_a|-\mb{p},\downarrow\rangle_b-|\mb{p},\downarrow\rangle_a|-\mb{p},\uparrow\rangle_b\Big]
\end{equation}
where $|\mb{p},\uparrow\rangle$ ($|\mb{p},\downarrow\rangle$) represents a sate with momentum $\mb{p}$ and spin pointing in the $\bs{\hat{z}}$ ($-\bs{\hat{z}}$) direction, with $\psi(\mb{p})$ given, in spherical coordinates $(r,\theta,\phi)$, by
\begin{equation}\label{wavefunction}
	\psi(p, \theta_p,\phi_p)\propto \delta(p-m_b\gamma_{v_b} v_b),
\end{equation}
$m_b$ being the mass of particle $b$, $v_b$ the modulus of its velocity  and $\gamma_{v_b}\equiv1/\sqrt{1-v_b^2}$. We are using a system of units in which the speed of light in vacuum is $c=1$. Considering the state of Eq. (\ref{singlet}) with the wavefunction of Eq. (\ref{wavefunction}), we see that each particle can go in any direction, but the particles propagate always in opposite directions and with the same absolute value for the momentum, in a singlet state of spin. This state can be obtained, for instance, through the decay of a spin-0 particle in two spin-1/2 particles in the center-of-mass rest frame of the system. The form of the state is a direct consequence of the conservation of momentum and angular momentum in the process. In this paper, we are following the Wigner's definition of spin \cite{wigner39}, that correspond to the angular momentum of the particle in its own rest frame. We will also consider here that the mass of particle $a$ is much greater than the mass of particle $b$, such that particle $a$ has non-relativistic velocities in the state of Eq. (\ref{singlet}). This last assumption is made only to maximize the relativistic effects that will be presented.

Note that if the spin and the momentum of relativistic particles could be treated as independent variables, the state of Eq. (\ref{singlet}) could be written as the product of a state for the particles momenta and a state for the particles spin. Upon tracing out the particles momenta, we would have a maximally entangled state for the particles spin. However, as we will discuss in this paper, it is not possible to treat the spin and the momentum of relativistic particles as independent variables, so we cannot trace out the particles momenta and define a reduced density matrix for the particles spin.

If we make joint measurements of spin in both particles considering eigenvalues $\pm1$ for each measurement, the CHSH version of Bell's inequalities states that for any local realistic description of the correlations among the particles we must have \cite{clauser69}
\begin{equation}\label{chsh}
	S=|\langle \hat{\mb{a}}_1,\hat{\mb{b}}_1\rangle+\langle \hat{\mb{a}}_1,\hat{\mb{b}}_2\rangle+\langle \hat{\mb{a}}_2,\hat{\mb{b}}_1\rangle-\langle \hat{\mb{a}}_2,\hat{\mb{b}}_2\rangle|\le2,
\end{equation}
where $\langle \hat{\mb{a}}_i,\hat{\mb{b}}_j\rangle\equiv\langle \mb{a}_i\cdot\hat{\bs{\sigma}}\otimes\mb{b}_j\cdot\hat{\bs{\sigma}}\rangle$ with $\mb{a}_i$ and $\mb{b}_j$ being unit vectors and $\hat{\bs{\sigma}}\equiv\hat{\sigma}_x\bs{\hat{x}}+\hat{\sigma}_y\bs{\hat{y}}+\hat{\sigma}_z\bs{\hat{z}}$, $\hat{\sigma}_x$, $\hat{\sigma}_y$ and $\hat{\sigma}_z$ being the spin-1/2 Pauli matrices. $\langle \hat{\mb{a}}_i,\hat{\mb{b}}_j\rangle$ represents the expectation value of a joint spin measurement with quantization axis in the direction $\mb{a}_i$ for particle $a$ and in the direction $\mb{b}_j$ for particle $b$. For the  sate of Eq. (\ref{singlet}) we have $\langle \hat{\mb{a}}_i,\hat{\mb{b}}_j\rangle=-\mb{a}_i\cdot{\mb{b}}_j$, such that for $\mb{a}_1=\bs{\hat{x}}$, $\mb{a}_2=\bs{\hat{y}}$, $\mb{b}_1=(\bs{\hat{x}}+\bs{\hat{y}})/\sqrt{2}$ and $\mb{b}_2=(\bs{\hat{x}}-\bs{\hat{y}})/\sqrt{2}$ we have $S=2\sqrt{2}\approx2.83>2$, indicating that quantum mechanics is not a local realistic theory. This fact makes entangled states like the one of Eq. (\ref{singlet}) very important in the field of quantum information. 

%$\mathcal{S}(0.99c,\theta,\phi)$ $\theta$ $\phi$

\begin{figure}\begin{center}
  \includegraphics[width=7.5cm]{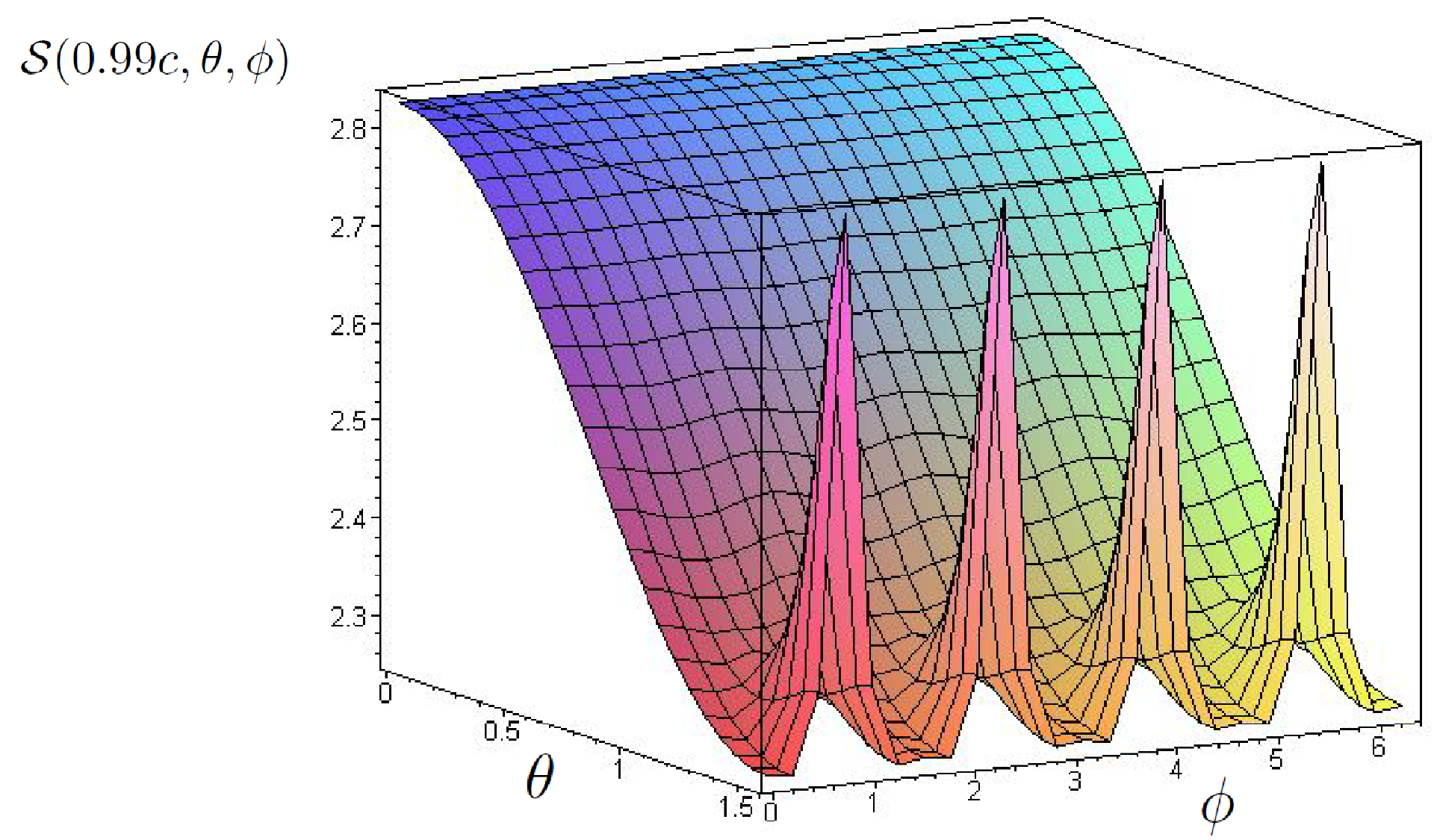}\\
  \caption{$\mathcal{S}(v_b,\theta,\phi)$, defined as $S$ from Eq. (\ref{chsh}) for spin measurements made  by Stern-Gerlach apparatuses with magnetic fields in the directions $\mb{B}_{a1}\propto\bs{\hat{x}}$, $\mb{B}_{a2}\propto\bs{\hat{y}}$, $\mb{B}_{b1}\propto(\bs{\hat{x}}+\bs{\hat{y}})/\sqrt{2}$ and $\mb{B}_{b2}\propto(\bs{\hat{x}}-\bs{\hat{y}})/\sqrt{2}$ for the state of Eq. (\ref{singlet}) with the  wavefunction of Eq. (\ref{wavefunction}) with $v_b=0.99c$ in function of the direction of the velocity of particle $b$ in spherical coordinates.}
\label{fig1}
 \end{center}\end{figure}

\begin{figure}\begin{center}
  \includegraphics[width=7.0cm]{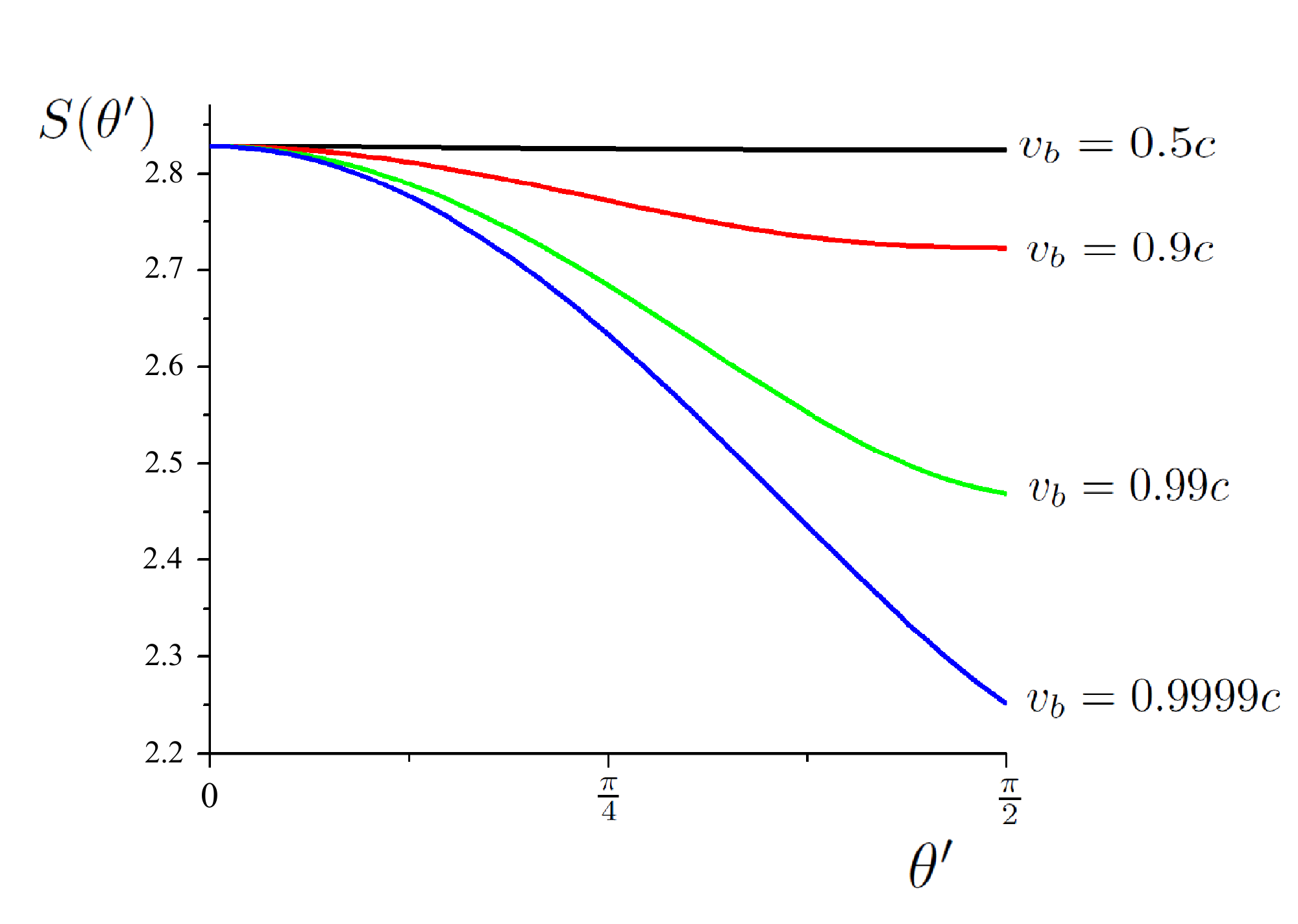}\\
  \caption{$S(\theta')$ from Eq. (\ref{chsh_av}) for spin measurements made by Stern-Gerlach apparatuses with magnetic fields in the directions $\mb{B}_{a1}\propto\bs{\hat{x}}$, $\mb{B}_{a2}\propto\bs{\hat{y}}$, $\mb{B}_{b1}\propto(\bs{\hat{x}}+\bs{\hat{y}})/\sqrt{2}$ and $\mb{B}_{b2}\propto(\bs{\hat{x}}-\bs{\hat{y}})/\sqrt{2}$ for the state of Eq. (\ref{singlet}) with the wavefunction of Eq. (\ref{wavefunction}) with $v_b=0.5c$, $v_b=0.9c$, $v_b=0.99c$ and $v_b=0.9999c$.}
\label{fig2}
 \end{center}\end{figure}

However, considering that particle $b$ has relativistic velocities $\mb{v}_b$ in the state of Eq. (\ref{singlet}), if we physically implement a spin measurement using an experimental apparatus of the Stern-Gerlach type, in which spin couples to an electromagnetic field, each velocity component of the particle will see a different quantization axis for the measurement. This occurs because, since spin refers to the particle angular momentum in its own rest frame, being proportional to the particle magnetic moment in the rest frame, the quantization axis of a spin measurement is in the direction of the apparatus magnetic field in the particle rest frame \cite{saldanha12}. But to compute the magnetic field in the particle rest frame from the magnetic field in the laboratory frame we must apply the corresponding Lorentz transformation, that depends on the particle velocity.  If the apparatus magnetic field is $\mb{B}$ in the laboratory frame (assumed here to be the same as the particles center-of-mass rest frame), in the particle $b$ rest frame it will be \cite{jackson} 
\begin{equation}
	\mb{B}_0=\gamma_{v_b} \mb{B}-\frac{\gamma_{vb}^2(\mb{v}_b\cdot\mb{B})}{\gamma_{vb}+1}\mb{v}_b,
\end{equation}
 with $\gamma_{v_b}\equiv1/\sqrt{1-v_b^2}$,  and the quantization axis of the measurement will be in the direction of $\mb{B}_0$. If we consider CHSH measurements with the magnetic fields of the apparatuses in the directions $\mb{B}_{a1}\propto\bs{\hat{x}}$, $\mb{B}_{a2}\propto\bs{\hat{y}}$, $\mb{B}_{b1}\propto(\bs{\hat{x}}+\bs{\hat{y}})/\sqrt{2}$ and $\mb{B}_{b2}\propto(\bs{\hat{x}}-\bs{\hat{y}})/\sqrt{2}$ in the laboratory frame, the amount of violation of the inequality of Eq. (\ref{chsh}) will depend on $\mb{v}_b$, since the quantization axes of the measurements depend on $\mb{v}_b$, and we can define a function $\mathcal{S}(v_b,\theta,\phi)$ that includes the dependence of the quantity $S$ from Eq. (\ref{chsh}) on the velocity of particle $b$ in spherical coordinates. Fig. \ref{fig1} illustrates  $\mathcal{S}(v_b,\theta,\phi)$ for $v_b=0.99c$ and $v_a$ non-relativistic. We see that the amount of violation of the Bell's inequality depend on the direction of propagation of the particle.  However, for each particle velocity we can choose other pair of directions for the magnetic fields $\mb{B}_{b1}$ and $\mb{B}_{b2}$ in the laboratory frame such that the Bell's inequality is maximally violated. But if we are making spin measurements with detectors that have an acceptance angle $\theta'$, making a post-selection of particles $b$ propagating in directions within cones  with $\theta<\theta'$, an average of the situations illustrated in the graph of Fig. \ref{fig1} will be the result of the measurement of the quantity $S$ from Eq. (\ref{chsh}):
\begin{equation}\label{chsh_av}
	S(\theta')=\frac{\int_0^{\theta'} d\theta \int_0^{2\pi} d\phi\; \sin(\theta)\mathcal{S}(v_b,\theta,\phi)}{\int_0^{\theta'} d\theta \int_0^{2\pi} d\phi\; \sin(\theta)},
\end{equation}
since each momentum component that enters in the detector in general will have different quantization axes for the spin measurements.
Fig. \ref{fig2} shows the values of the quantity $S(\theta')$ from Eq. (\ref{chsh_av}) for different modulus of the velocities $v_b$ in the wavefunction of Eq. (\ref{wavefunction}). The smaller is the acceptance angle of the detector $\theta'$, higher is the value of $S(\theta')$, but smaller is the fraction of particles that are detected.

Fig. \ref{fig2} shows that the amount of violation of Bell's inequalities for the state of Eq. (\ref{singlet}) depends on the velocity distribution of the particles, even the reduced spin density matrix of this state being maximally entangled. Only with a post-selection of particles $b$ in momentum eigenstates ($\theta'\rightarrow0$ in Fig. \ref{fig2}) we obtain a maximal violation. This example illustrates why the association of a reduced spin density matrix for relativistic particles and the quantification of the entanglement of this reduced state, as frequently done in the literature \cite{gingrich02,li03,lamata06,jordan07,landulfo09,friis10,choi11}, is meaningless. We cannot predict the expectation values of measurements without considering the velocities of the particles. It may be argued that this fact occurs because we choose Stern-Gerlach apparatuses to implement the measurements, and that it may exists other kinds of apparatuses that measure the particle spin independently of the particle velocity, as conjectured by Peres \textit{et al.} \cite{peres02}. But a spin measurement must be made through the coupling of the 3 components of spin with some 3-component quantity of the measuring apparatus, represented by a scalar interaction Hamiltonian. In a covariant treatment, the interaction Hamiltonian must be proportional to an invariant scalar to guarantee that all observers in inertial reference frames compute the same expectation values for the spin measurements \cite{saldanha12}. This fact imply that spin must transform under Lorentz transformation in the same way as the physical quantity to which it couples in the measuring apparatus. So the quantity that couples with spin in the measuring apparatus computed in the particle rest frame will, in any case, depend on the particle velocity, such that it is not possible to measure the particle spin independently from its velocity (or momentum), no matter how we measure spin. 

%Suppose now that we have a quantum state that is a statistical mixture of the state of Eq. (\ref{singlet}) with $v_b=0.9999c$ and a state $\rho'$ with the same momentum distribution but a identity matrix for spin: $\rho=p|\Psi\rangle\langle\Psi|+(1-p)\rho'$ with $0\leq p\leq1$. According to Fig. \ref{fig2}, for the state $|\Psi\rangle$, $S_\Psi(\pi/2)$ from Eq. (\ref{chsh_av}) is approximately $2.27$. For the state $\rho'$ we have $S_{\rho'}(\pi/2)=0$. So, for the state $\rho$ we have $S_\rho(\pi/2)\approx 2.27p$. For $p<0.9$ it is not possible to violate the CHSH inequality, but the state has spin entanglement for $p>x$. So it is possible to have an entangled spin state that do not present non-classical spin correlations unless we make a post-selection on the particles velocities. 

If the particles center-of-mass propagates with velocity $\bs{\beta}=\beta\bs{\hat{z}}$ in relation to the laboratory frame, the state viewed in the laboratory frame would be the  state of Eq. (\ref{singlet}) transformed by a Lorentz boost with velocity $-\bs{\beta}$. However, to see what would be the amount of violation of the Bell's inequality of Eq. (\ref{chsh}) in this situation, it is easier to compute the direction of the apparatuses magnetic fields in each particle rest frame.  Since the directions of the magnetic fields of the Stern-Gerlach apparatuses in the laboratory frame to be used are orthogonal to the center-of-mass velocity, in the center-of-mass rest frame the electromagnetic  fields are \cite{jackson} $\mb{B}'=\gamma_\beta\mb{B}$ and  $\mb{E}'=\gamma_\beta\bs{\beta}\times\mb{B}$ with $\gamma_\beta\equiv1/\sqrt{1-\beta^2}$, $\mb{B}$ being the magnetic field in the laboratory frame and $\mb{B}'$ and $\mb{E}'$ being the magnetic and electric fields in the particles center-of-mass frame. Since particle $a$ is assumed to have a nonrelativisitc velocity in relation to the particles center-of-mass, the quantization axis of each measurement for this particle is in the same direction as the magnetic field in the laboratory frame. However, since particle $b$ is assumed to have relativistic velocities $\mb{v}_b$ in relation to the particles center-of-mass, the quantization axis for each velocity component is in the direction of the vector \cite{jackson}
\begin{equation}\label{transf_B}	\mb{B}_0'=\gamma_{v_b}\gamma_\beta[\mb{B}-\mb{v}_b\times(\bs{\beta}\times\mb{B})]-\frac{\gamma_\beta\gamma_{vb}^2}{\gamma_{vb}+1}\mb{v}_b(\mb{v}_b\cdot\mb{B}),
\end{equation}
that represents the magnetic field in the particle rest frame.

\begin{figure}\begin{center}
  \includegraphics[width=7.0cm]{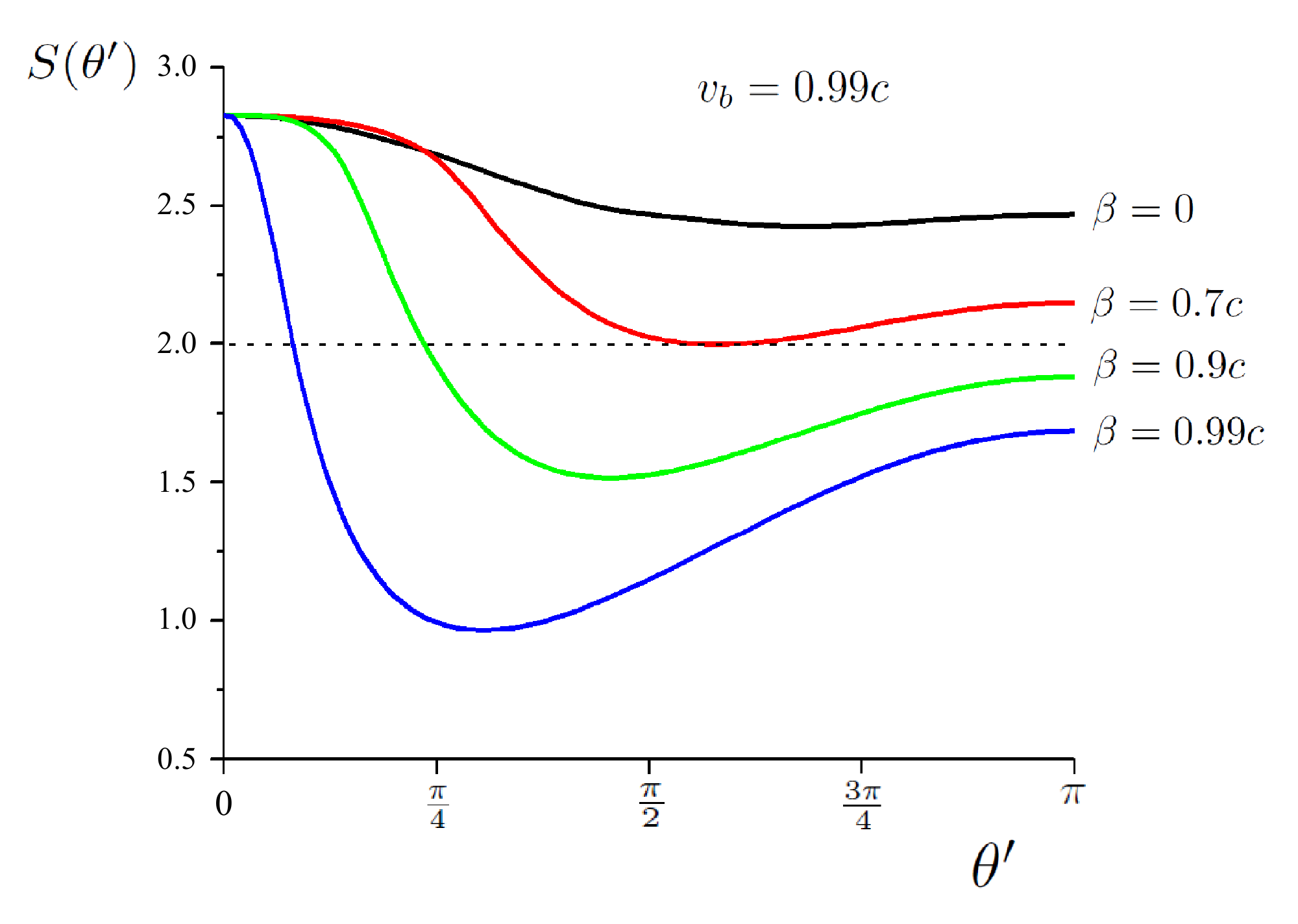}\\
  \caption{$S(\theta')$ from Eq. (\ref{chsh_av}) for spin measurements made by Stern-Gerlach apparatuses with magnetic fields in the directions $\mb{B}_{a1}\propto\bs{\hat{x}}$, $\mb{B}_{a2}\propto\bs{\hat{y}}$, $\mb{B}_{b1}\propto(\bs{\hat{x}}+\bs{\hat{y}})/\sqrt{2}$ and $\mb{B}_{b2}\propto(\bs{\hat{x}}-\bs{\hat{y}})/\sqrt{2}$ for the state of Eq. (\ref{singlet}) with the wavefunction of Eq. (\ref{wavefunction}) with  $v_b=0.99c$  in a reference frame that moves with velocity $\beta\bs{\hat{z}}$ in relation to the center-of-mass rest frame, with $\beta=0$, $\beta=0.7c$, $\beta=0.9c$ and $\beta=0.99c$.}
\label{fig3}
 \end{center}\end{figure}

We can compute the amount of violation of Bell's inequality in Eq. (\ref{chsh}) considering the above dependence of the quantization axis of the spin measurement of particle $b$ with $\mb{v}_b$, $\bs{\beta}$ and $\mb{B}$.  Choosing again $\mb{B}_{a1}\propto\bs{\hat{x}}$, $\mb{B}_{a2}\propto\bs{\hat{y}}$, $\mb{B}_{b1}\propto(\bs{\hat{x}}+\bs{\hat{y}})/\sqrt{2}$ and $\mb{B}_{b2}\propto(\bs{\hat{x}}-\bs{\hat{y}})/\sqrt{2}$, considering the state from Eq. (\ref{singlet}) in the particles center-of-mass frame with the wavefunction of Eq. (\ref{wavefunction}) with $v_b=0.99c$, we plot in Fig. \ref{fig3} the values for $S(\theta')$ from Eq. (\ref{chsh_av}), noting that now $\mathcal{S}$ also depends on $\bs{\beta}$, for different velocities $\bs{\beta}$ between the particles center-of-mass and the laboratory frame. Note that a post-selection of particles $b$ propagating in directions within cones with $\theta<\theta'$ in relation to the particles center-of-mass correspond to different acceptance angles $\theta''$ of the detectors for each velocity $\bs{\beta}$. We opted for this representation in the graphics of Fig. \ref{fig3} because in this way we are always post-selecting the same portion of the state in each situation. We can see that the amount of violation of Bell's inequality is different in different reference frames, and that in some frames we cannot even violate it without a post-selection of the sate.

Let us consider now another situation in which Alice prepares the state of Eq. (\ref{singlet}) in her laboratory, with the particles rest frame coinciding with the laboratory frame, and send particle $b$ with a selected velocity $\mb{v}_b$ to Bob. Consider that Alice's reference frame is moving with a relativistic velocity $\bs{\beta}=\beta\bs{\hat{z}}$ in relation to Bob's frame. If they want to maximally violate a Bell's inequality with a set of shared pair of particles obtained in this way, Bob has to adjust his Stern-Gerlach apparatus with magnetic fields in directions $\mb{B}$ such that the directions of the vectors $\mb{B}_0'$ from Eq. (\ref{transf_B}) are the directions that, together with the quantization axes chosen by Alice, maximally violate Eq. (\ref{chsh}).

In summary, we had shown how the dependence of the quantization axis of a spin measurement on the particle velocity  influences the violation of the CHSH version of Bell's inequalities with relativistic spin-1/2 particles in different situations. Our treatment takes into account how spin measurements can be physically implemented, making different predictions in relation to the previous works on the subject. Even if the particles are in a maximally entangled spin state, we may not maximally violate the Bell's inequalities without a post-selection of the particles momenta, such that the definition of a reduced density matrix for the particles spin is meaningless. The spin and the momentum of relativistic particles cannot be considered independent variables in a relativistic treatment. We also showed that observers in different reference frames measure different amounts of violation for the CHSH inequality, and that in some reference frames the inequality may not be violated. We believe that many of our predictions can be experimentally tested with an apparatus similar to the one from Sakai \textit{et al.} experiments \cite{sakai06}.

P.L.S. acknowledges Nicolai Friis for useful discussions and the Brazilian agencies CAPES, CNPq and FACEPE for the financial support. V.V. acknowledges financial support from the National Research Foundation and Ministry of Education in Singapore and the support of Wolfson College Oxford.

%$\theta\;\;$ $\phi$ $S\;\;$  $S(\theta')\;\;$ $\theta'\;\;$ $\phi'\;\;$ $\frac{\pi}{2}\;\;$ $\frac{\pi}{4}\;\;$ $\frac{3\pi}{4}\;\;$  $\pi\;\;$

%$v_b=0.5c$ $v_b=0.9c$ $v_b=0.99c$ $v_b=0.9999c$

%$\mathcal{S}(0.99c,\theta,\phi)$

%$\beta=0$ $\beta=0.9c$ $\beta=0.99c$ $\beta=0.9999c$

%\bibliography{references}

\begin{thebibliography}{30}
\expandafter\ifx\csname natexlab\endcsname\relax\def\natexlab#1{#1}\fi
\expandafter\ifx\csname bibnamefont\endcsname\relax
  \def\bibnamefont#1{#1}\fi
\expandafter\ifx\csname bibfnamefont\endcsname\relax
  \def\bibfnamefont#1{#1}\fi
\expandafter\ifx\csname citenamefont\endcsname\relax
  \def\citenamefont#1{#1}\fi
\expandafter\ifx\csname url\endcsname\relax
  \def\url#1{\texttt{#1}}\fi
\expandafter\ifx\csname urlprefix\endcsname\relax\def\urlprefix{URL }\fi
\providecommand{\bibinfo}[2]{#2}
\providecommand{\eprint}[2][]{\url{#2}}



\bibitem{bell}
\bibinfo{author}{\bibfnamefont{J. S.}~\bibnamefont{Bell}},
  \emph{\bibinfo{title}{Speakable and Unspeakable in Quantum Mechanics}} 
    (\bibinfo{publisher}{Cambridge University Press}, \bibinfo{address}{Cambridge}, \bibinfo{year}{1987}).


\bibitem[{\citenamefont{Peres}(2002)}]{peres}
\bibinfo{author}{\bibfnamefont{A.}~\bibnamefont{Peres}},
  \emph{\bibinfo{title}{Quantum Theory: Concepts and Methods}}
 (\bibinfo{publisher}{Kluwer Academic Publishers}, \bibinfo{address}{New
  York}, \bibinfo{year}{2002}).


\bibitem[{\citenamefont{Clauser et~al.}(1969)\citenamefont{Clauser, Horne,
  Shimony, and Holt}}]{clauser69}
\bibinfo{author}{\bibfnamefont{J.~F.} \bibnamefont{Clauser}},
  \bibinfo{author}{\bibfnamefont{M.~A.} \bibnamefont{Horne}},
  \bibinfo{author}{\bibfnamefont{A.}~\bibnamefont{Shimony}}, \bibnamefont{and}
  \bibinfo{author}{\bibfnamefont{R.~A.} \bibnamefont{Holt}},
  \bibinfo{journal}{Phys. Rev. Lett.} \textbf{\bibinfo{volume}{23}},
  \bibinfo{pages}{880} (\bibinfo{year}{1969}).

\bibitem[{\citenamefont{Czachor}(1997)}]{czachor97}
\bibinfo{author}{\bibfnamefont{M.}~\bibnamefont{Czachor}},
  \bibinfo{journal}{Phys. Rev. A} \textbf{\bibinfo{volume}{55}},
  \bibinfo{pages}{72} (\bibinfo{year}{1997}).


\bibitem[{\citenamefont{Peres et~al.}(2002)\citenamefont{Peres, Scudo, and
  Terno}}]{peres02}
\bibinfo{author}{\bibfnamefont{A.}~\bibnamefont{Peres}},
  \bibinfo{author}{\bibfnamefont{P.~F.} \bibnamefont{Scudo}}, \bibnamefont{and}
  \bibinfo{author}{\bibfnamefont{D.~R.} \bibnamefont{Terno}},
  \bibinfo{journal}{Phys. Rev. Lett.} \textbf{\bibinfo{volume}{88}},
  \bibinfo{pages}{230402} (\bibinfo{year}{2002}).

\bibitem[{\citenamefont{Peres and Terno}(2004)}]{peres04}
\bibinfo{author}{\bibfnamefont{A.}~\bibnamefont{Peres}} \bibnamefont{and}
  \bibinfo{author}{\bibfnamefont{D.~R.} \bibnamefont{Terno}},
  \bibinfo{journal}{Rev. Mod. Phys.} \textbf{\bibinfo{volume}{76}},
  \bibinfo{pages}{93–123} (\bibinfo{year}{2004}).

\bibitem[{\citenamefont{Alsing and Milburn}(2002)}]{alsing02}
\bibinfo{author}{\bibfnamefont{P.~M.} \bibnamefont{Alsing}} \bibnamefont{and}
  \bibinfo{author}{\bibfnamefont{G.~J.} \bibnamefont{Milburn}},
  \bibinfo{journal}{Quantum Inf. Comput.} \textbf{\bibinfo{volume}{2}},
  \bibinfo{pages}{487} (\bibinfo{year}{2002}).

\bibitem[{\citenamefont{Gingrich and Adami}(2002)}]{gingrich02}
\bibinfo{author}{\bibfnamefont{R.~M.} \bibnamefont{Gingrich}} \bibnamefont{and}
  \bibinfo{author}{\bibfnamefont{C.}~\bibnamefont{Adami}},
  \bibinfo{journal}{Phys. Rev. Lett.} \textbf{\bibinfo{volume}{89}},
  \bibinfo{pages}{270402} (\bibinfo{year}{2002}).

\bibitem[{\citenamefont{Ahn et~al.}(2003)\citenamefont{Ahn, Lee, Moon, and
  Hwang}}]{ahn03}
\bibinfo{author}{\bibfnamefont{D.}~\bibnamefont{Ahn}},
  \bibinfo{author}{\bibfnamefont{H.~J.} \bibnamefont{Lee}},
  \bibinfo{author}{\bibfnamefont{Y.~H.} \bibnamefont{Moon}}, \bibnamefont{and}
  \bibinfo{author}{\bibfnamefont{S.~W.} \bibnamefont{Hwang}},
  \bibinfo{journal}{Phys. Rev. A} \textbf{\bibinfo{volume}{67}},
  \bibinfo{pages}{012103} (\bibinfo{year}{2003}).

\bibitem[{\citenamefont{Li and Du}(2003)}]{li03}
\bibinfo{author}{\bibfnamefont{H.}~\bibnamefont{Li}} \bibnamefont{and}
  \bibinfo{author}{\bibfnamefont{J.}~\bibnamefont{Du}}, \bibinfo{journal}{Phys.
  Rev. A} \textbf{\bibinfo{volume}{68}}, \bibinfo{pages}{022108}
  (\bibinfo{year}{2003}).

\bibitem[{\citenamefont{Terashima and Ueda}(2003)}]{terashima03}
\bibinfo{author}{\bibfnamefont{H.}~\bibnamefont{Terashima}} \bibnamefont{and}
  \bibinfo{author}{\bibfnamefont{M.}~\bibnamefont{Ueda}},
  \bibinfo{journal}{Quantum Inf. Comput.} \textbf{\bibinfo{volume}{3}},
  \bibinfo{pages}{224} (\bibinfo{year}{2003}).

\bibitem[{\citenamefont{Lee and Chang-Young}(2004)}]{lee04}
\bibinfo{author}{\bibfnamefont{D.}~\bibnamefont{Lee}} \bibnamefont{and}
  \bibinfo{author}{\bibfnamefont{E.}~\bibnamefont{Chang-Young}},
  \bibinfo{journal}{New J. Phys.} \textbf{\bibinfo{volume}{6}},
  \bibinfo{pages}{67} (\bibinfo{year}{2004}).

\bibitem[{\citenamefont{Kim and Son}(2005)}]{kim05}
\bibinfo{author}{\bibfnamefont{W.~T.} \bibnamefont{Kim}} \bibnamefont{and}
  \bibinfo{author}{\bibfnamefont{E.~J.} \bibnamefont{Son}},
  \bibinfo{journal}{Phys. Rev. A} \textbf{\bibinfo{volume}{71}},
  \bibinfo{pages}{014102} (\bibinfo{year}{2005}).

\bibitem[{\citenamefont{Caban and Rembieli\'nski}(2005)}]{caban05}
\bibinfo{author}{\bibfnamefont{P.}~\bibnamefont{Caban}} \bibnamefont{and}
  \bibinfo{author}{\bibfnamefont{J.}~\bibnamefont{Rembieli\'nski}},
  \bibinfo{journal}{Phys. Rev. A} \textbf{\bibinfo{volume}{72}},
  \bibinfo{pages}{012103} (\bibinfo{year}{2005}).

\bibitem[{\citenamefont{Lamata et~al.}(2006)\citenamefont{Lamata,
  Martin-Delgado, and Solano}}]{lamata06}
\bibinfo{author}{\bibfnamefont{L.}~\bibnamefont{Lamata}},
  \bibinfo{author}{\bibfnamefont{M.~A.} \bibnamefont{Martin-Delgado}},
  \bibnamefont{and} \bibinfo{author}{\bibfnamefont{E.}~\bibnamefont{Solano}},
  \bibinfo{journal}{Phys. Rev. Lett.} \textbf{\bibinfo{volume}{97}},
  \bibinfo{pages}{250502} (\bibinfo{year}{2006}).

\bibitem[{\citenamefont{Jordan et~al.}(2007)\citenamefont{Jordan, Shaji, and
  Sudarshan}}]{jordan07}
\bibinfo{author}{\bibfnamefont{T.~F.} \bibnamefont{Jordan}},
  \bibinfo{author}{\bibfnamefont{A.}~\bibnamefont{Shaji}}, \bibnamefont{and}
  \bibinfo{author}{\bibfnamefont{E.~C.~G.} \bibnamefont{Sudarshan}},
  \bibinfo{journal}{Phys. Rev. A} \textbf{\bibinfo{volume}{75}},
  \bibinfo{pages}{022101} (\bibinfo{year}{2007}).

\bibitem[{\citenamefont{Landulfo and Matsas}(2009)}]{landulfo09}
\bibinfo{author}{\bibfnamefont{A.~G.~S.} \bibnamefont{Landulfo}}
  \bibnamefont{and} \bibinfo{author}{\bibfnamefont{G.~E.~A.}
  \bibnamefont{Matsas}}, \bibinfo{journal}{Phys. Rev. A}
  \textbf{\bibinfo{volume}{79}}, \bibinfo{pages}{044103}
  (\bibinfo{year}{2009}).

\bibitem[{\citenamefont{Caban et~al.}(2010)\citenamefont{Caban, Dziegielewska,
  Karmazyn, and Okrasa}}]{caban10}
\bibinfo{author}{\bibfnamefont{P.}~\bibnamefont{Caban}},
  \bibinfo{author}{\bibfnamefont{A.}~\bibnamefont{Dziegielewska}},
  \bibinfo{author}{\bibfnamefont{A.}~\bibnamefont{Karmazyn}}, \bibnamefont{and}
  \bibinfo{author}{\bibfnamefont{M.}~\bibnamefont{Okrasa}},
  \bibinfo{journal}{Phys. Rev. A} \textbf{\bibinfo{volume}{81}},
  \bibinfo{pages}{032112} (\bibinfo{year}{2010}).

\bibitem[{\citenamefont{Friis et~al.}(2010)\citenamefont{Friis, Bertlmann,
  Huber, and Hiesmayr}}]{friis10}
\bibinfo{author}{\bibfnamefont{N.}~\bibnamefont{Friis}},
  \bibinfo{author}{\bibfnamefont{R.~A.} \bibnamefont{Bertlmann}},
  \bibinfo{author}{\bibfnamefont{M.}~\bibnamefont{Huber}}, \bibnamefont{and}
  \bibinfo{author}{\bibfnamefont{B.~C.} \bibnamefont{Hiesmayr}},
  \bibinfo{journal}{Phys. Rev. A} \textbf{\bibinfo{volume}{81}},
  \bibinfo{pages}{042114} (\bibinfo{year}{2010}).



\bibitem[{\citenamefont{Choi et~al.}(2011)\citenamefont{Choi, Hur, and
  Kim}}]{choi11}
\bibinfo{author}{\bibfnamefont{T.}~\bibnamefont{Choi}},
  \bibinfo{author}{\bibfnamefont{J.}~\bibnamefont{Hur}}, \bibnamefont{and}
  \bibinfo{author}{\bibfnamefont{J.}~\bibnamefont{Kim}},
  \bibinfo{journal}{Phys. Rev. A} \textbf{\bibinfo{volume}{84}},
  \bibinfo{pages}{012334} (\bibinfo{year}{2011}).

\bibitem[{\citenamefont{Wigner}(1939)}]{wigner39}
\bibinfo{author}{\bibfnamefont{E.}~\bibnamefont{Wigner}},
  \bibinfo{journal}{Ann. Math.} \textbf{\bibinfo{volume}{40}},
  \bibinfo{pages}{149} (\bibinfo{year}{1939}).

\bibitem[{\citenamefont{Weinberg}(1995)}]{weinberg}
\bibinfo{author}{\bibfnamefont{S.}~\bibnamefont{Weinberg}},
  \emph{\bibinfo{title}{The Quantum Theory of Fields}},
  vol.~\bibinfo{volume}{1} (\bibinfo{publisher}{Cambridge University Press},
  \bibinfo{address}{Cambridge}, \bibinfo{year}{1995}).

\bibitem[{\citenamefont{Saldanha and Vedral}(2011)}]{saldanha12}
\bibinfo{author}{\bibfnamefont{P.~L.} \bibnamefont{Saldanha}} \bibnamefont{and}
  \bibinfo{author}{\bibfnamefont{V.}~\bibnamefont{Vedral}}, New J. Phys. \textbf{14}, 023041 (2012).

\bibitem[{\citenamefont{Jackson}(1999)}]{jackson}
\bibinfo{author}{\bibfnamefont{J.~D.} \bibnamefont{Jackson}},
  \emph{\bibinfo{title}{Classical Electrodynamics}} (\bibinfo{publisher}{John
  Wiley \& Sons}, \bibinfo{address}{New York}, \bibinfo{year}{1999}),
  \bibinfo{edition}{3rd} ed.

\bibitem[{\citenamefont{\textit{et al.}}(2006)}]{sakai06}
\bibinfo{author}{\bibfnamefont{H.~Sakai} \bibnamefont{\textit{et al.}}},
  \bibinfo{journal}{Phys. Rev. Lett.} \textbf{\bibinfo{volume}{97}},
  \bibinfo{pages}{150405} (\bibinfo{year}{2006}).

\end{thebibliography}

\end{document}